\documentclass[floats,prb,twocolumn]{revtex4}

\usepackage{graphicx}
\usepackage{amsfonts}
\usepackage{amssymb}
\usepackage{amsmath}

\newcommand{\qnr}[1]{\lfloor #1 \rfloor_{q}}

\def\der#1#2{{\partial#1 \over \partial#2}}
\def\be{\begin{equation}}
\def\ee{\end{equation}}
\def\ba#1{\begin{array}{#1}}
\def\ea{\end{array}}
\def\bn{\begin{enumerate}}
\def\en{\end{enumerate}}

\def\rr{\right}
\def\l{\left}

\def\H{\mathcal{H}}
\def\Q{\mathcal{Q}}
\def\P{\mathcal{P}}
\def\summ{\sum\limits}

\def\He{H_{\text{eff}}}

\begin{document}

\title{Permutation Symmetric Critical Phases in Disordered Non-Abelian Anyonic Chains }
\author{L. Fidkowski}
\author{G. Refael}
\author{H-H. Lin}
\affiliation{Department of Physics, Institute for Quantum Information, California Institute of
  Technology, MC 114-36, Pasadena, CA 91125}
\author{P. Titum}
\affiliation{Indian Institute of Technology Kanpur, India}

%\pacs{}
%\date{\today}
\begin{abstract}
Topological phases supporting non-abelian anyonic excitations have been proposed as candidates for topological quantum computation.  In this paper, we study disordered non-abelian anyonic chains based on the quantum groups $SU(2)_k$, a hierarchy that includes the $\nu=5/2$ FQH state and the proposed $\nu=12/5$ Fibonacci state, among others.  We find that for odd $k$ these anyonic chains realize infinite randomness critical {\it phases} in the same universality class as the $S_k$ permutation symmetric multi-critical points of Damle and Huse (Phys. Rev. Lett. 89, 277203 (2002)).  Indeed, we show that the pertinent subspace of these anyonic chains actually sits inside the ${\mathbb Z}_k \subset S_k$ symmetric sector of the Damle-Huse model, and this ${\mathbb Z}_k$ symmetry stabilizes the phase.
\end{abstract}

\maketitle

\section{Introduction}

One of the major advances in the understanding of strongly correlated 
quantum systems has been the
exploration of topological phases of matter.  Originating with the
discovery of the $\nu=1/3$ fractional quantum Hall effect, topological
phases have received much renewed interest with their recent proposed
application to quantum computation \cite{kitaev}.  Under this proposal, quantum computation is carried out by the braiding of the non-abelian quasi-particle excitations of the topological phase.  The topologically protected degenerate space of ground states of the non-abelions serves as the memory, and the braiding induces unitary transformations within this Hilbert space \cite{collins}.
The remarkable feature of this scheme is that the dimension of this space for $N$ anyons grows as $d^N$, with the quantum dimension $d$ in general non-integer.  This should be contrasted, for example, with $N$ spin-1/2 quasi-particles, whose Hilbert space has $2^N$ states.  The non-integral nature of $d$ reflects a unique non-locality of the anyon Hilbert space and makes decoherence-free quantum computation possible: no local perturbations can give rise to decoherence.

The potential applications of topological phases with non-abelian
anyons, and perhaps even more so their remarkable properties, warrant an
analysis of interacting many-anyon systems.  Possibly the most natural
starting point is a $1$-dimensional
anyonic chain, analogous to regular spin chains such as the $SU(2)$
Heisenberg model.  Though such anyonic chains do not have truly local
degrees of freedom, they can be written as local systems with local
constraints.  It was found that even rather simple to construct
translationally invariant chains, based on $SU(2)_3$ anyons, have an
intricate structure, with the low energy degrees of freedom organized
into the gapless spectra of the $c=4/5$ and $c=7/10$ conformal field
theories \cite{Qpeople1}.  These chains interact via nearest neighbor
couplings and are exactly solvable, though more generalized models
exhibit an even richer set of critical and gapped phases \cite{Qpeople2}.

Whereas translationally invariant chains are the requisite first step, a more likely physical realization of a non-abelian chain, for example in a quantum Hall state, will have a strong degree of quenched
disorder.  Such chains are the focus of this paper. Even with
garden-variety spins, the addition of disorder dramatically affects
the physics, with the low energy behavior controlled by so-called
infinite randomness fixed points \cite{DSF94, DSF95}.  A prototypical
example is the random singlet phase describing the state of a
disordered spin-$1/2$ Heisenberg chain.  Here spins pair up and form
singlets in a random fashion, with most connecting near neighbors but
some being very long ranged.  This unique structure of the ground
state leads to some rather unexpected or unusual properties, including
algebraically decaying average correlations and energy-length scaling
$|\ln E| \sim L^{\psi}$ as opposed to the usual $1/E\sim L^{z}$
characteristic of pure systems. Furthermore, disordered spin-$S$ chains
with $S>1/2$ were shown to exhibit infinite-randomness critical fixed points with
the universal exponent $\psi=1/(2S)$, and with the spin-state described
as a competition between $2S+1$ domains, as we describe below. These
fixed points were dubbed the $S_n$ permutation symmetric
points, with $n=2S+1$ the number of competing domains \cite{DamleHuse, MGJ,HymanYang,RKF}.

Because of the unique structure of their Hilbert spaces, disordered
anyonic chains are particularly amenable to treatment via strong
randomness renormalization group methods, and indeed have been shown
to exhibit infinite randomness fixed points \cite{YB, us}.  They are
thus an especially fertile ground for trying to discover and classify
new universality classes of strongly random behavior.  Indeed, even
though no new universality classes were found, Ref. [\onlinecite{us}] found
a rich phase diagram for the $SU(2)_3$, or Fibonacci anyonic chain,
with a random singlet fixed point that can be destabilized by the
addition of couplings favoring fusion into a non-trivial topological
charge, and a resulting flow to a more intricate $S_3$ permutation
symmetric fixed point.  It is notable that this $S_3$ symmetric point
is actually a critical phase, with no relevant perturbations - this
contrasts with the domain realization of the $S_n$ symmetric fixed
point, which has $n-1$ relevant perturbations \cite{DamleHuse}. 
 
The non-abelian models that are the subject of this paper are based on
$SU(2)_k$ anyons with $k$ odd.  The level $k$ signifies a truncation of all representations with spin $S > k/2$ - the anyons of $SU(2)_k$ simply correspond to the first $k+1$ irreducible representations of $SU(2)$.  This fact potentially suggests a subtle analogy between the $SU(2)_k$ systems and spins $S$ truncated at $k/2$.  In
particular, it raises the possibility of a relation between the $S_{k}$ permutation symmetric fixed points of regular spin-$(k-1)/2$ chains, and the infinite randomness phases arising in $SU(2)_k$ anyons. 

In this paper we show that, indeed, for all odd $k$, $SU(2)_k$ anyonic chains
realize $S_k$ symmetric infinite randomness critical {\it phases}.
Crucial to our analysis is expressing the $2$-anyon interaction terms
in a novel basis, one which behaves better than the standard projector
basis with respect to the renormalization group (RG) decimations.  With this insight, we are
able to solve the model, and in fact construct an explicit equivalence
between it and the ${\mathbb D}_k \subset S_k$ symmetric sector of the
domain model of Damle and Huse \cite{DamleHuse}, where the order $2k$
dihedral group ${\mathbb D}_k$ in particular contains ${\mathbb Z}_k$.
The ${\mathbb Z}_k$ symmetry is what stabilizes the phase, eliminating
the $k-1$ relevant perturbations of the domain model.  This phenomenon
of a $k$ multi-critical point being stabilized by additional symmetry
in the $SU(2)_k$ model is tantalizingly close to that discovered in
[\onlinecite{Qpeople3}] for the uniform $SU(2)_k$ chain.  There,
stability of an a priori $k$ multi-critical point is guaranteed by an
extra "topological" symmetry in the quantum system.  It is conceivable
that these two phenomena are closely related.

The rest of this paper is structured as follows.  In section II we
briefly review necessary background on the strong randomness
renormalization group procedure, as well as set up the construction of
the anyon chain Hilbert space and Hamiltonian.  In Section III we
analyze the disordered anyon chain, introducing the novel basis for
the interaction terms.  We write down and solve the flow equations,
finding a fixed point of the RG.  In Section IV we construct an
explicit equivalence between the disordered anyon chain and the
${\mathbb D}_k$ symmetric sector of Damle Huse domain wall model, one
that relates their respective fixed points.  We conclude in Section V,
and relegate some technical derivations to the appendix.

\section{Background and Setup}

\subsection{Real Space Renormalization Group}

To find the ground state of disordered spin chains, Ma and
Dasgupta introduced the strong disorder real-space renormalization
group method \cite{MaDas1979,MaDas1980}.  The random spin-$1/2$
Heisenberg model provides the simplest example amenable to such a treatment.
The model is given by: \be H=\sum_i J_{i, i+1} \, {\bf S}_i \cdot
{\bf S}_{i+1}, \ee where the couplings $J_{i,i+1}>0$ are positive
and randomly distributed.  Note that, as far as the Hilbert space
is concerned, we have for two neighboring sites \be \label{tp}
\frac{1}{2} \otimes \frac{1}{2} = 0 \oplus 1, \ee and the
interaction ${\bf S}_i \cdot {\bf S}_{i+1}$ simply gives an energy splitting
between the singlet and triplet.  The procedure now is to pick the largest $J_{i,i+1}$, which
effectively decimates the excited triplet and leaves the ground
state in a singlet, and do perturbation theory around that state.
Quantum fluctuations then induce an effective coupling according
to the so-called Ma-Dasgupta rule \cite{MaDas1979, MaDas1980}: \be
J_{i-1,i+2} = \frac{J_{i-1,i} J_{i+1,i+2}}{2 J_{i,i+1}} \label{md}
\ee So sites $i$ and $i+1$ are decimated and replaced with an
effective interaction between $i-1$ and $i+2$.  Iteration of this
procedure produces singlet bonds on all length scales.  This is the random
singlet ground state.

A quantitative description is obtained by tracking the RG flow of the
coupling distribution.  It is useful to employ logarithmic couplings
\cite{DSF94}:
\be
\beta_{i,i+1} = \ln \frac{\Omega}{J_{i,i+1}}
\ee
where $\Omega = \text{max}_i \, J_{i,i+1}$.  In these variables the
Ma-Dasgupta rule (\ref{md}) reads
\be
\beta_{i-1,i+2} = \beta_{i-1,i} +
\beta_{i+1,i+2}
\ee
(up to an additive constant of $\ln 2$, which can be safely neglected).  As the couplings get decimated $\Omega$
decreases.  It is convenient to define the RG flow parameter as
\be
\Gamma = \ln \frac{\Omega_0}{\Omega},
\ee
where $\Omega_0$ is the maximal coupling of the bare Hamiltonian.  Let
$P_\Gamma(\beta)$ be the distribution of couplings.  We can derive a
flow equation for $P_\Gamma(\beta)$ by decimating the couplings in the
infinitesimal interval $\beta = [\,0, d \Gamma \,]$ and seeing how
their probabilistic weight is redistributed.  We obtain
\be
\ba{c}
\frac{d}{d \Gamma} P_\Gamma (\beta) = \der{P_\Gamma}{\beta} +  \vspace{2mm}\\
P(0) \int_0^\infty d \beta_1 \int_0^\infty d \beta_2
  \delta(\beta-\beta_1-\beta)  P_\Gamma(\beta_1)  P_\Gamma(\beta_2)
\ea
\ee
The first term comes from the overall
change of scale, and the second from the Ma-Dasgupta rule.  These
equations have a solution
\be
P_\Gamma (\beta) = \frac{1}{\Gamma} e^{-\beta / \Gamma}
\ee
which is an attractive fixed point to essentially all physical initial
configurations.  This solution permits us to read off features of the
random singlet phase; for example one can with a little more work
derive the energy-length scaling relation:
\be
L^{1/2} \sim \Gamma = \ln \, (1/E).
\ee
which thus has the exponent:
\be
\psi=1/2.
\ee

\begin{figure}
\includegraphics[width=7.5cm]{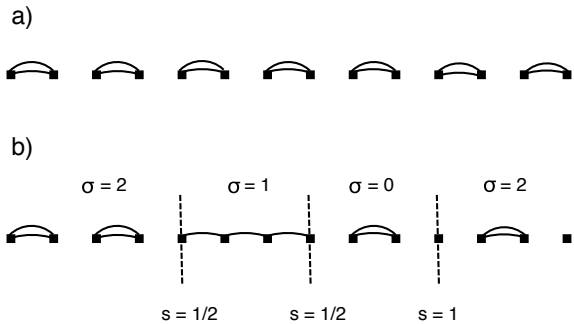}
\caption{a) Valence Bond Solid of type $\sigma=2$.  b) Sample
  configuration at the infinite randomness fixed point.  All the
  domains occur equally.  The dashed lines represent domain walls,
  which contain the unpaired spins that form the low energy degrees of
  freedom. \label{acfvb}}
\end{figure}

The random singlet phase describes the universal low-energy behavior
of several known one dimensional systems, making it interesting to
attempt to classify all such universal low energy fixed points of
strongly random systems in one dimension.  Thus far all known
universality classes are realized in the Damle-Huse hierarchy of
permutation symmetric critical points \cite{DamleHuse}, indexed by a
positive integer $n$.  In the construction of Damle and Huse, the
system indexed by $n$ is realized in a spin-$n/2$ $SU(2)$ invariant
Heisenberg model, with the sites represented as symmetrized tensor
products of spin $1/2$'s.  The interaction Hamiltonian is
\be 
H = \sum_{i} J_{i} \, \vec{\bf S}_i \cdot \vec{\bf S}_{i+1} 
\ee 
where
the couplings $J_i$ contain dimerization $\delta$ and randomness of
strength $R$ \be J_i = J \left[ 1 + \delta (-1)^i \right] \exp (R
\eta_i). 
\ee 
Here the $\eta_i$ are random variables.  Depending on $R$ and $\delta$
this Hamiltonian can realize a plethora of phases, which can be
qualitatively understood in terms of Valence Bond Solids (VBS).  In
this picture, a VBS of type $\sigma \in \{0,1,\ldots,n \}$ is
constructed by pairing up $\sigma$ spin-$1/2$'s into singlets over
each even bond and $n-\sigma$ spin-$1/2$'s over each odd bond (figure
\ref{acfvb}. a).  This exhausts the spin degrees of freedom and
defines a unique state.

The $S_n$ permutation symmetric infinite randomness fixed point is now realized as a
multi-critical point in which all of these domains occur equally
(figure \ref{acfvb}. b).  The domain walls contain unpaired spins:
between $\sigma$ and $\sigma'$ we have a spin of magnitude $|\sigma -
\sigma'| / 2$.  These spins interact via effective couplings whose
magnitude is highly dependent on the domain that separates them, and
whose sign is dictated by a consistency condition.  Specifically, for
three domains $\sigma_1, \sigma_2$, and $\sigma_3$, the interaction
between $|\sigma_1 - \sigma_2| / 2$ and $|\sigma_2 - \sigma_3| / 2$ is
anti-ferromagnetic if $\sigma_1 - \sigma_2$ and $\sigma_3 - \sigma_2$
are of the same sign and ferromagnetic otherwise.  The fixed point
turns out to contain an entirely random distribution of domains,
described by a stochastic transfer matrix with all nonzero entries
equal to $1/n$, an energy-length infinite randomness scaling exponent
$\psi=1/n$, and a logarithmic distribution of the coupling
strengths: $P(\beta) = (n/\Gamma) \, e^{-n \beta / \Gamma}$.  

We will see in the remainder of the paper that this multi-critical
point will be realized (for $n+1$ odd) as a stable phase of the
$SU(2)_k$ anyon chain.  First, however, we need to review some
background on anyonic spin chains.

\subsection{Hilbert Space and Hamiltonian of $SU(2)_k$ Anyon Chains \label{HSNA}}

A crucial part of our analysis relies on the specific properties of $SU(2)_k$.  The nontrivial
anyons in this case correspond to the nontrivial representations of
$SU(2)$ at level $k$.  There are $k$ of these, labeled by their spin:
$0, 1/2, 1, \ldots, k/2$.  The fusion rules are:
\be \label{fusion_eq}
i \otimes j = \sum_{m= |i-j|}^{\text{min } (i+j, k-i-j)} m,
\ee
where $m$ is summing over the integers if $i-j$ is an integer, and
over the half integers otherwise.  More information is encoded in the so-called $F$-matrix or set of
$q$-$6$-$j$ symbols of $SU(2)_k$.  Given $3$ anyons $j_1,j_2$, and
$j_3$, we can either fuse $j_1$ and $j_2$ first into $j_{12}$ and then
with $j_3$ into $j$, or we could first fuse $j_2$ and $j_3$ into
$j_{23}$ and then with $j_1$ into $j$.  Both of these procedures
generate a basis for the the Hilbert space of ground states of $j_1$,
$j_2$, and $j_3$.  The transformation between these two bases is
encoded in the $F$-matrix $( {F \,}_{j_1 j_2}^{j_3 j}
)_{j_{12}}^{j_{23}}$.  The $F$-matrix of $SU(2)_k$ is written down in
the appendix.

Let us now construct the Hilbert space for the problem.  We have a
chain $\{ i_p \}$ of anyons indexed by an integer position $p$.  From
now on we will deal only with odd level $k$ and integer values of the
"spin" - thus $i_p \in \{ 1, \ldots, (k-1)/2 \}$.  We can do this because the integers form a closed fusion sub-algebra of $SU(2)_k$.  Indeed, there is a ${\mathbb Z}_2$ symmetry that relates charge $j$ and $k/2-j$, and this symmetry will become important when we relate the anyon model to the Damle-Huse model.

\begin{figure}
\includegraphics[width=5.5cm]{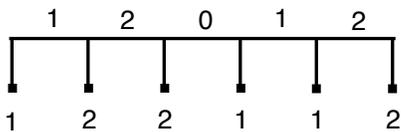}
\caption{Sample link basis vector for an $SU(2)_5$ chain.  The anyons are represented by the black boxes at the bottom and labeled with their topological charge, $1$ or $2$.  The links, or bonds, between the anyons are labeled with $0$, $1$, or $2$, in such a way that the fusion rules at the trivalent vertices are obeyed. \label{acf1}}
\end{figure}

There are two equivalent ways to define the Hilbert space \cite{us}.  The simplest construction is to label each "link" between site $p$ and $p+1$ with an integer anyon type $l_{p,p+1}$, subject to the
constraint that the fusion rules be obeyed at each site, i.e.,
\be 
l_{p,p+1} \in \{|l_{p-1,p} - i_p|, \ldots, \text{min }(l_{p-1,p} +
i_p, k - l_{p-1,p} - i_p) \} 
\ee 
See, for example, figure \ref{acf1}.  The set of all such admissible
labelings defines a basis for the Hilbert space, which is simply the
space of degenerate ground states of this configuration of anyons.

We can impose the link basis constraints as high energy $2$-body
interactions.  The advantage of the link basis is then that it gives
us a local way to describe the degrees of freedom in the problem.
Indeed, we will see that the Hamiltonian defined below will consist of
$3$-local interactions.  In order to define the Hamiltonian, however,
it is useful to first describe a second, more abstract way to define
the Hilbert space.

\begin{figure}
\includegraphics[width=6.5cm]{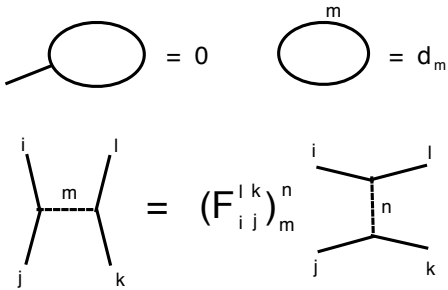}
\caption{$F$-matrix reconnection rules.  The first is the so-called "no tadpole" rule, stating that any graph that can be disconnected by the removal of one edge is $0$.  The second states that removing a loop with label $m$ is equivalent to multiplying by the quantum dimension $d_m = ({\bf F}_{mm}^{mm})_0^0$.  The third states that by performing the indicated reconnection on any local portion of the graph, we have the stated linear relation among the resulting graphs. \label{acf2}}
\end{figure}

Let us for convenience suppose the chain is finite, consisting of $N$
anyons.  We first consider the space of all trivalent graphs, with
endpoints on the $N$ anyons, whose edges are labeled by the
non-trivial integer anyon types, and whose vertices satisfy the fusion
rules.  We take the Hilbert space generated by such graphs (modulo
graph isomorphism) and quotient out the subspace generated by the
$F$-matrix relations, interpreted as local reconnection rules (see
figure \ref{acf2}).  To relate this graphical picture to the link
basis, note that the link basis states can be viewed as labeled
trivalent graphs, and that any other labeled trivalent graph can be
reduced to a superposition of these using $F$-matrix reconnection
rules.  The inner product of two graphs is defined by reflecting one
of the graphs and concatenating it with the other along the $N$ nodes.

\begin{figure}
\includegraphics[width=6.5cm]{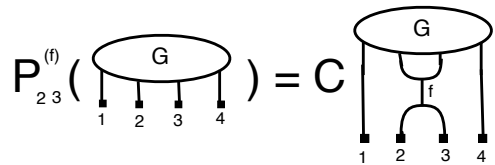}
\caption{Graphical representation of a projector onto total topological charge $f$ acting on sites $2$ and $3$.  The oval labeled by $G$ represents a graph as described in the graphical definition of the Hilbert space.  The normalization constant $C$ is chosen so as to make ${\P}$ a projector. \label{acf3}}
\end{figure}

An advantage of this graphical picture of the Hilbert space is that it makes it easy to describe the interaction terms occurring in the Hamiltonian.  The Hamiltonian is a sum over $p$ of pairwise interactions between sites $p$ and $p+1$.  These are constrained by the $SU(2)_k$ symmetry, and therefore a linear combination of projection operators onto some total topological charge $f$:

\be
\H=\summ_{p,f} J_{p, p+1}^f {\P}_{p,p+1}^{(f)}
\label{hanyon}
\ee These projection operators ${\P}$ have a graphical representation, and in the abstract graph basis their action on a particular graph $G$ is simply given by concatenation of ${\P}$ with $G$, up to a normalization constant (see figure \ref{acf3}).  Their action in the link basis can be worked out by concatenating ${\P}$ with a particular link graph and then using $F$-matrix rules to reduce the resulting graph to a linear combination of link graphs.  In this manner it is apparent that the action of ${\P}_{p,p+1}$ depends only on $l_{p-1,p}$, $l_{p,p+1}$, and $l_{p+1,p+2}$; it is thus a $3$-local operator.

\section{Analysis of the Disordered $SU(2)_k$ chain}

\subsection{Convenient Basis for the RG}

We would like to apply the real space RG procedure to the disordered
$SU(2)_k$ chain (\ref{hanyon}) in hopes of finding infinite randomness
fixed points.  Let us first review what happened in our previous
analysis [\onlinecite{us}] of the case $k=3$, i.e., the Fibonacci
chain.  $SU(2)_3$ contains one nontrivial anyon of integer charge, the
so-called $\tau$ anyon, and the chain is simply an array of these.  In
the strong randomness limit, we applied the Ma-Dasgupta rule to the
strongest bond, which was a projection operator on a pair of
neighboring $\tau$'s.  These $\tau$'s could fuse to one of two
possible states, either one with trivial total topological charge or
another $\tau$ ($\tau\otimes\tau=0\oplus\tau$), and the bond projected onto one of these, leading to
either the elimination of both anyons, or their merger into one.  In
either case, we were left with effectively another realization of the
Fibonacci chain, with either $1$ or $2$ fewer sites, allowing us to
iterate the procedure.

For $SU(2)_k$ a new complication arises.  This time, when we pick the
largest bond to decimate, the generic situation is that there are more
than two possible fusion products for the corresponding pair of
anyons.  For example, in $SU(2)_5$, the fusion rules are: 
\be
\ba{c} 
1 \otimes 1 =
0 \oplus 1 \oplus 2,\\
2\otimes 2=0\oplus 1,\\
2\otimes 1=1\oplus 2.
\ea\ee 
The first rule, regarding the spin $1$ representation, contains $3$
possible fusion products on the right hand side.  We would like to be
able to keep just the lowest energy of these, in order to merge the
two anyons into a single new effective anyonic site.  In
general, however, we are not allowed to do this: when there are more than two
fusion products, there will be more than one energy splitting
associated with the bond.  While we can decimate away the largest one,
there may be couplings on other bonds that need to be decimated before
the smaller couplings on the original strong bond.  This decimation of
only one fusion product leads to a situation where on the one hand we
need to enforce a constraint on the two anyons, but on the other we
are not allowed to merge them into a single effective anyon.  This
impasse makes it seemingly impossible to carry out an iterative real
space RG analysis. \footnote{This can be
circumvented, however, by renormalizing the topological charges such
that a certain fusion channel is eliminated. For instance, if we want
to exclude the $1\otimes 1\rightarrow 2$ channel between two sites, we
can demote the two sites from charge 1 to 2, with the fusion channels
now possible $2\otimes 2=0\oplus 1$ (this follows the procedure first
suggested in \cite{MGJ}). In the limit of strong disorder, however,
this is not necessary.}

\begin{figure}
\includegraphics[width=5.5cm]{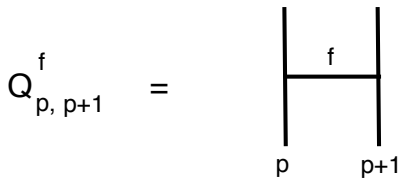}
\caption{Graphical definition of the operator ${\Q}_{p,p+1}^{(f)}$.  Note that it is nonzero only when the fusion rules at the $(i_p, i_p, f)$ and $(i_{p+1},i_{p+1}, f)$ vertices are obeyed. \label{acf4}}
\end{figure}

In the rest of this paper we demonstrate the existence and stability
of an infinite randomness fixed point for $SU(2)_k$ chains that
circumvents the above difficulty.  The idea is to construct a
Hamiltonian out of two-site operators ${\Q}_{p,p+1}$
\be 
\H = \sum_p J_{p,p+1} {\Q}_{p,p+1} 
\ee 
such that $\H$ retains its
form under an RG procedure where we truncate all excited fusion
products.  In other words, the effective operators generated from first and second order decimations are all proportional to $\Q$.  The {\it a priori} assumption that all such excited states
can be truncated, which is not valid in general, is then justified in
this particular case if we can show that the resulting RG leads to a
strongly disordered set of couplings $J_{p,p+1}$.  This is because
when we express the operator ${\Q}_{p,p+1}$ as a linear combination of
projection operators
\be 
{\Q}_{p,p+1} = \sum_f c_f {\P}_{p,p+1}^{(f)}, 
\ee 
the differences
between the coefficients $c_f$ remain of order $1$, independently of
the broadness of the distribution of $\log J_{p,p+1}$.  Hence the
energy splittings for each bond are all of the same order, and with
very high probability all get decimated in one fell swoop in the large
disorder limit.  The assumption that all excited states can be
truncated is then justified and the scheme is self-consistent.  Of course, this does not rule out the possiblity of more exotic phases where the interactions are not built out of only the $\Q$ operators - these phases, if they existed, would not be amenable to treatment by this method.

We now claim that the correct operators to use are the ones defined
graphically in figure \ref{acf4} and denoted ${\Q}^{(f)}$.  They can
be thought of as an exchange of an anyon of topological charge $f$;
their action on any graph $G$ is simply by concatenation with $G$, as
for any graphically defined operator.  The ${\Q}^{(f)}$ differ from
the projection operators ${\P}_{p,p+1}^{(g)}$ by an $F$-matrix move,
and can of course be expanded as linear combinations of the
${\P}_{p,p+1}^{(g)}$.  They all have well defined scaling under the
RG, and, as we show in the appendix and discuss in more detail below,
the most relevant one is ${\Q}^{(1)}$.  Thus from now on we will
consider a Hamiltonian of the form
\be 
\H=\summ_p J_{p, p+1} {\Q}_{p,p+1}^{(1)} \label{hanyon1} 
\ee 
where the couplings $J_{p, p+1}$ are disordered.  In the appendix, we
show 
\be \label{apeq}
{\Q}^{(1)} = \sum_{f=|i - j|}^{\text{min }(i+j,k-i-j)} A(f)
\, {\P}^{(f)} 
\ee
where 
\be 
A(f)= \qnr{f}^2 + \qnr{f+1}^2 - \qnr{|i -
  j|}^2 - \qnr{i+j + 1}^2 \label{q11} 
\ee 
is an increasing function of
$f$.  Here $i$ and $j$ are the topological charges of the two
neighboring anyons on which ${\Q}^{(1)}$ acts, and the $q$-numbers are
defined as 
\be \label{qnreq}
\qnr{n} = \frac{q^n - q^{-n}}{q-q^{-1}} 
\ee 
with $q=e^{\pi i / (k+2)}$.

Thus, depending on the sign of the coupling $J_{p,p+1}$, the anyons at
$p$ and $p+1$ can fuse to either an anyon of charge $| i_p - i_{p+1}
|$ or one of charge $\text{min } (i_p + i_{p+1}, k - i_p - i_{p+1})$.
As discussed above, we decimate all the other fusion products.  This
scheme will be self consistent if we show that the system flows to
strong disorder, i.e., that the distribution of the $\log J_{p,p+1}$
broadens out.  To analyze the flow, we first need to work out the
decimation rules.

\begin{figure}
\includegraphics[width=7cm]{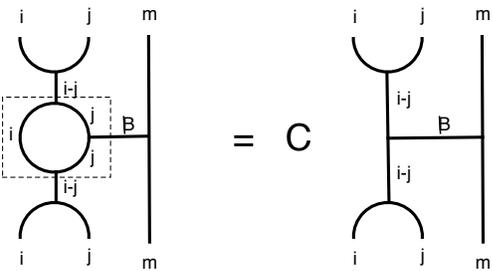}
\caption{Graphical representation of the effective operator between the composite $i-j$ and its neighbor $m$, generated at first order in perturbation theory.  It is equal to ${\P}_{12}^{(i-j)} {\Q}_{23}^{(\beta)} {\P}_{12}^{(i-j)}$.   By performing $F$-matrix manipulations within the dashed box, we see that this effective operator is just a multiple of ${\Q}_{\text{comp. } 3}^{(\beta)}$. \label{acf5}}
\end{figure}

First, let us focus on first order decimations.  Here the strongest
bond $J_{p,p+1}$ fuses the $p$ and $p+1$ anyons into a composite, and
the effective interactions between the composite and its neighbors
depend on the original $<p-1,p>$ and $<p+1,p+2>$ interactions.  Now,
for the Hamiltonian (\ref{hanyon1}) we see from figure \ref{acf5} that
these effective interactions are proportional to ${\Q}_{p-1,p}$ and
${\Q}_{p+1,p+2}$ - the figure represents graphically the first order
perturbation calculation \cite{us}.  In fact, we can say more: figure
\ref{acf5} makes it clear that not only are the ${\Q}^{(1)}$
preserved, but so are all the ${\Q}^{(\beta)}$.  More precisely, if we
think of a first order decimation as a linear mapping from the space
of interactions between say $p+1$ and $p+2$ (or $p$ and $p-1$) to the
space of interactions between the composite and $p+2$ (or $p-1$),
figure \ref{acf5} makes it clear that the operators
${\Q}_{p,p+1}^{(\beta)}$ are eigenvectors of this mapping.

Second order decimations follow a similar paradigm, and in the
appendix we show that, again, the ${\Q}^{(\beta)}$ are eigenvectors of
the second order decimations, and derive the real-space Ma-Dasgupta
decimation step, which (for $\Q^{(1)}$) reads:
\be
\H_{eff}= C_j \, \frac{J_{1,2}J_{3,4}}{J_{2,3}} \, \Q^{(1)}_{1,4},
\ee
as in Eq. (\ref{2ndDEC}), with $C_j$ given by the expression (\ref{cj}).  For both first and second order
decimations, we also calculate in the appendix the corresponding
eigenvalues for all $\beta$, and show that in each case they are maximized for
$\beta=1$.  Thus ${\Q}^{(1)}$ is the most relevant operator, which is
why we chose to construct the Hamiltonian out of it in
(\ref{hanyon1}).  It is stable with respect to perturbations by the
${\Q}_{p,p+1}^{(\beta)}$ for $\beta \geq 2$, which, having smaller
eigenvalues under decimation, are irrelevant.

In fact, the stability argument is not entirely rigorous, but follows
the usual line of justification for the validity of strong randomness
RG.  Basically one can show that the addition of a small amount of
${\Q}_{p,p+1}^{(\beta)}$ for $\beta>1$ doesn't change the decimation
rules, and so these ${\Q}_{p,p+1}^{(\beta)}$ decay away under the RG,
up to bad spots, or "cancers", that occur with frequency that vanishes
with increasing randomness.  We can invoke the standard line of
reasoning used to justify strong randomness RG in the first place to
argue that these do not destabilize the fixed point, though ultimately
this should be decided by numerical simulation.

Thus, with the ansatz (\ref{hanyon1}) for the Hamiltonian, we have a
consistent framework for the RG that eliminates the potential
multitude of widely distributed energy scales associated to each bond.
Instead, we have only one energy scale for each bond, multiplying the
operator ${\Q}_{p,p+1}^{(1)}$.  In the strong randomness limit,
decimation of the strongest bond results, with probability approaching
$1$, in the decimation of {\it all} its excited states, leaving a $0$
or $1$ anyons in place of $2$.  This decimation process preserves the
form of the interactions ${\Q}_{p,p+1}^{(1)}$.  All that we have left
to do is to show that under the RG the Hamiltonian (\ref{hanyon1})
does indeed flow to strong randomess.  This we do in the next
subsection.

\vskip 1pt
\subsection{Flow Equations}

Let us now solve the model (\ref{hanyon1}).  To do this, we need to
derive the flow equations that describe real space RG.  We are dealing
with an ensemble of chains, where not only the coupling strengths and
signs, but also the anyon types are chosen according to some
probability distribution (see also [\onlinecite{MGJ}]).  We make the ansatz that the coupling
strengths, signs, and anyon types are completely uncorrelated from
each other, and uncorrelated among the different sites/bonds.  Also,
for simplicity we analyze only the case in which the bond strength
probability distribution is symmetric with respect to sign change.
The stability analysis in this framework might in principle miss
asymmetric relevant perturbations, and indeed other non-independent
distribution perturbations, but the exact mapping to (a ${\mathbb
  Z}_k$ symmetric subspace of) the Damle-Huse model discussed in the
next subsection will show that there are none, at least at the level
of analysis in [\onlinecite{DamleHuse}].

Let $R(i), i=1,\ldots,(k-1)/2$ be the probability distribution for the
(integer) anyon types, and $P(\beta)$ the logarithmic bond-strength
probability distribution, normalized to integrate out to $1/2$ because
of the two possibilities for the sign of the coupling.  Here $\beta =
\log (\Omega / J)$ is the logarithmic coupling, and $\Omega$ the
energy cutoff.  Considering both first and second order decimations,
we obtain the following infinitesimal transformation for the joint
probability distribution $R(i) \, P(\beta)$:

\begin{widetext}
\be \label{eqdec}
R(i) \, P(\beta) \rightarrow R(i) \, P(\beta+d\Gamma) + d\Gamma \,P(0)
\, (R \otimes R) (i) P(\beta) + d\Gamma \, P(0) \, (R \otimes R)(0) \,R(i) \, (P\otimes P)(\beta) 
\ee 
The notation is defined below.  Here the first term comes from the cutoff rescaling, the second from first order decimations, and the third from second order decimations.  Equation (\ref{eqdec}) is equivalent to the
following two norm preserving transformations of $R(i)$ and $P(\beta)$:

\be
\ba{c}R(i) \rightarrow R(i) + P(0) \, d\Gamma \, (R\otimes R)(i) -P(0) \, d\Gamma \, \left( 1 - (R \otimes R)(0) \right) \, R(i) \\
P(\beta) \rightarrow P(\beta) + d\Gamma P'(\beta) + P(0) \, d\Gamma \,
(R \otimes R)(0) \, (P \otimes P)(\beta) - P(0) \, d\Gamma \,
\left(1-(R \otimes R)(0) \right) P(\beta)
\ea
\ee
\end{widetext} 
The notation is as follows.  We define the convolution \be (P \otimes
P) (\beta) = 2 \int_0^\beta d\beta' \, P(\beta') \, P(\beta - \beta')
\ee with an extra factor of $2$ to account for normalization of $P$,
and we let \be (R \otimes R) (i) = \sum_{j,l=1}^{(k-1)/2} \eta(i,j,l)
\, R(j) \, R(l) \ee where $\eta(i,j,l)$ is equal to $1$ if $i = |j-l|$
or $i = \text{min }(j+l, k-j-l)$ and is $0$ otherwise (the two
possibilities correspond to the two possible signs of the coupling). 

We can also write down the integro-differential flow equations
corresponding to these infinitesimal transformations (the $\Gamma$
dependence is implicit):
\begin{widetext}
\be
\ba{c}
\frac{dR(i)}{d\Gamma} = P(0) \, (R\otimes R)(i) - P(0) \, \left(1-(R \otimes R)(0) \right) \, R(i) \\
\frac{dP(\beta)}{d\Gamma} = P'(\beta) + 2 P(0) \, (R \otimes R) (0) \, \int_0^\beta d\beta' \, P(\beta') \, P(\beta - \beta') - P(0) \, \left(1-(R \otimes R)(0) \right) P(\beta)
\ea
\ee
\end{widetext} 
A solution to these equations is
\be
\ba{c} 
R_0(i) = \frac{2}{k-1} \nonumber \\ P_0(\beta)=\frac{k-1}{2 \Gamma}
e^{-(k-1) \beta / \Gamma} \label{soln} 
\ea
\ee 

Let us analyze the stability of the solution in Eq. (\ref{soln}).  First let us look at $R(i)$.  Consider a perturbation of the form $R(i) = R_0(i) + \epsilon_i$.  Using the easily derived fact that to linear order $(R \otimes R)(i) = (R_0 \otimes R_0)(i) - 2 \epsilon_i$ and the fact that $\sum_i \epsilon_i = 0$ we get the RG flow of $\epsilon_i$:

\be \frac{d \epsilon_i}{d \Gamma} = -P(0) \, \epsilon_i \ee which, since $P(0)$ is always positive, shows that $\epsilon_i$ always decays.  Now consider $P(\beta)$.  At linear order the variation in $(R \otimes R)(0)$ vanishes, so the analysis of the stability of $P(\beta)$ is as in all the other examples of strong randomness RG where this solution occurs \cite{DSF94, DSF95}.  As mentioned before, we don't consider perturbations asymmetric with respect to the sign of the coupling, but we expect these to be stable as well by arguments similar to those in [\onlinecite{us}].  We also show now via explicit mapping to the Damle-Huse model that they indeed are stable.

\section{Relation to the Damle-Huse Domain Wall Model}

As mentioned above, the fixed points we found above must somehow be related
to the Damle-Huse fixed points of abelian spin chains. In this section
we present and discuss the mapping between the Damle-Huse domain model
with $n=k$ domains and spin $S=(k-1)/2$, and our
$SU(2)_k$ anyonic chains, and show that this mapping gives an
equivalence between the permutation symmetric multi-critical point of
the domain model and the fixed point of the non-abelian anyonic chain.  In
addition, we show that the anyonic fixed point is actually a stable phase.

\begin{center}
\begin{figure*}
\includegraphics[width=15cm]{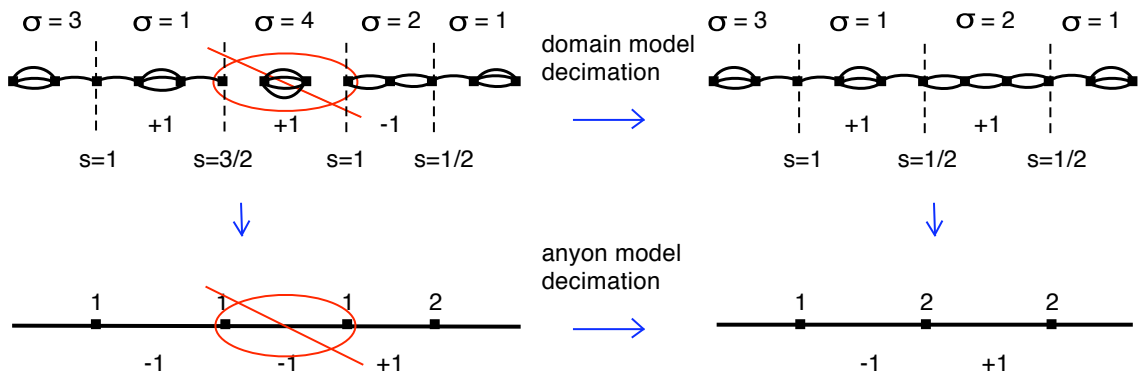}
\caption{An example illustrating commutativity of the equivalence mapping between the spin-$2$ Damle-Huse chain and the $SU(2)_5$ anyon chain and (first order) decimations.  A spin-$2$ configuration has five different domains, while the anyon
model has two non-trivial charges: $1$ and $2$. The domain model configuration at the upper left, has
  domains labeled by $\sigma = 3,1,4,2,1$, signs of 
  couplings $+1,+1,-1$, and strongest bond spanning the $4$
  domain. These are mapped to the $SU(2)_5$ domain (down arrow), with topological
  charges following (\ref{fmap}) as
  $f(\frac{|\sigma_{m+1}-\sigma_m|}{2})$, and signs of the couplings following (\ref{fmap2}),
  $s'(m)=(-1)^{(\sigma_{m+1}-\sigma_{m-1})}$. Decimating bond 4 (right
  arrow) in the anyon chain follows the principle stated in the appendix: when two anyons $i$ and $j$ fuse
ferromagnetically to $i+j<k/2$, the effective couplings are the same
as the original couplings (as opposed to other cases where the sign
may flip). Starting again from the top left corner, carrying out
 a real space decimation (right arrow) in the Damle-Huse chain first, and then the
  mapping to the anyon model (down arrow) results in the same configuration, as illustrated.
That this is true in general requires one to check a few more similar cases.  We thus see that the mapping from the spin-$(k-1)/2$ Damle-Huse chain to the $SU(2)_k$ chain and real-space RG steps commute.} \label{acfdec}
\end{figure*}
\end{center}

\subsection{Mapping Between the Two Models}

\begin{figure}
\includegraphics[width=8.5cm]{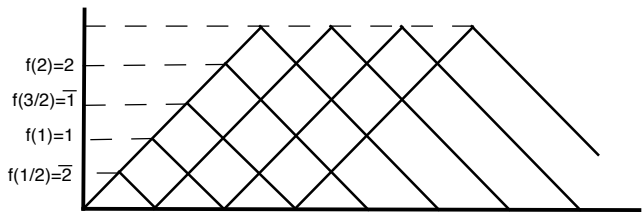}
\caption{Bratelli diagram for $SU(2)_5$.  The mapping $f$ defined below simply reflects the topological charges along the vertical axis for non-integer arguments, mapping non-integer charges to integers.  Bars denote the cases where the argument was non-integer and hence the action of $f$ non-trivial.  \label{bratt}}
\end{figure}

Some intuition for the mapping between the spin-$S$ Damle-Huse chain with $S=(k-1)/2$ and
the $SU(2)_k$ anyon chain can be obtained by inspecting the Bratteli diagram in
Fig. \ref{bratt}. Naively, we like to think of the possible
topological charges $i=0,\,1/2,\,\ldots,\,k/2$ of the $SU(2)_k$ tensor
category as somehow related to spin-$i$ representations of $SU(2)$. This
naive notion is not quite correct, because of the special constraints
that the Hilbert space truncation presents. As it turns out, both $i$
and $k/2-i$ essentially represent the same non-trivial topological
charge. Therefore the distinct non-trivial topological charges can be indexed
by integer $i$'s: $i=1,\,2,\,\ldots,\,k-1/2$; the half-integer
values of $i$ can be turned into integers through $i\rightarrow
k/2-i$. This is, for example, the reason for our ability to restrict our
rendition of the $SU(2)_5$ fusion algebra to the rules in (\ref{fusion_eq}) using only the charges 1 and 2, alongside the trivial (vacuum) charge $0$. 

The association of $i=0,\,1/2,\,\ldots,\,k/2$ with spin-$i$ $SU(2)$ representations provides the
correct intuition for the mapping between the anyon model and the spin $(k-1)/2$ Damle Huse domain model. In the spin-$(k-1)/2$ Damle-Huse chain, each site appears as a domain wall between two domains,
say $\sigma_m$ and $\sigma_{m+1}$ - in order to maintain transparent notation we label such a site by the pair $(m,m+1)$.  The spin of site $(m, m+1)$ is expected to be 
\be
S_{m,m+1}=\frac{|\sigma_{m+1}-\sigma_m|}{2}.
\ee
The range of possible spins is $1/2\le S_{m,m+1} \le (k-1)/2$, just like $i$ above, excluding the
two trivial (vacuum) charges $i=0,\,k/2$.  Thus in the mapping to the Damle Huse model, the natural thing to do is to identify the domain wall spins with the anyon charges.  However, we would also like to restrict to the integer topological charges.  This is naturally done through the mapping:
\be 
f(S)=\l\{\ba{cc}
S & S\in {\mathbb Z}\\
\frac{k}{2}-S & S+\frac{1}{2}\in{\mathbb Z}
\ea
\rr.
\label{fmap}
\ee 
with $2S \in {\mathbb Z}$, an integer, and $0< S<k/2$.
So for example, for $k=5$, we have
$f(2)=f(1/2) = 2$, and $f(1) = f(3/2) = 1$.  

Now let us define the mapping a little more formally.  A configuration in the Damle-Huse model is completely specified by a
sequence $\{ \sigma_m, \beta_m \}$ of domains $\sigma_m \in
{0,1,\ldots,k-1}$ and log couplings $\beta_m$ between them, since the
signs $s(m) = \pm 1$ of the couplings are uniquely determined by this
data.  As described above, we map this configuration to the sequence $\{ i_{(m,m+1)} =
f(|\sigma_{m+1} - \sigma_{m}| / 2) \}$ of $SU(2)_k$ anyons, with the
log of the coupling $J_m$ between anyons $i_{(m-1,m)}$ and $i_{(m,m+1)}$ given by
$\beta_m$, and the sign of $J_m$ given by  
\be
s'(m) = (-1)^{(\sigma_{m+1}-\sigma_{m-1})}s(m).
\label{fmap2}
\ee
This choice of sign will reproduce the prefered fusion channels
upon mapping a spin-$(k-1)/2$ chain (in its domain wall
representation) to an anyonic problem.

This mapping commutes, by construction, with both the first and second
order real-space RG decimations.  Since this claim is the key point in the proof,
for clarity, we illustrate it with a specific example in figure \ref{acfdec}.  In the caption we explain how we get the same anyon configuration and couplings, including signs, irrespective of whether we do the
decimation in the domain model and then map to the anyon model, or
first map to the anyon model and do the decimation there.  This means
that the two operations commute.  Second order decimations (not illustrated) are even easier to handle.  Here, in both the domain model and the anyon model, the sign of the coupling is given by
the Ma-Dasgupta rule (\ref{md}), and hence commute.  We have thus
defined a mapping from the configuration space of the Damle-Huse model
to that of our anyon model, and this mapping respects the RG
evolution.   It is quite remarkable that despite the two different origins of the interaction couplings' signs (one from the Damle-Huse domain model rules, and one from an $F$-matrix calculation), the signs conspire to
make the two operations commute. This is presumably a deep reflection of
the fact that the $SU(2)_k$ tensor categories were constructed from
the spin-representation of $SU(2)$. Also, note that while the
Ma-Dasgupta rules in the two models may have different multiplicative
constants, leading to a small difference in the logarithmic couplings
between the two models, this difference is unimportant in the large
disorder limit.

\subsection{Symmetry Considerations and Elimination of Relevant Perturbations}
 
Let us examine the properties of this mapping of configuration spaces.
First of all, we claim the map is onto, i.e., given any configuration
$\{ i_{(m-1,m)}, \beta_m, s'(m) \}$, there's a domain configuration $\{
\sigma_m, \beta_m, s(m) \}$ that maps onto it.  To see this, we first
pick any $0 \leq \sigma_1 \leq k-1$.  Then we must pick $0 \leq
\sigma_2 \leq k-1$ such that $f(|\sigma_2-\sigma_1|/2) = i_{(1,2)}$.  It
is easy to see that there are precisely $2$ choices of such $\sigma_2$
(naively we may think there are $4$, given the $2$ to $1$ nature of
both $f$ and the absolute value mapping, but $2$ of those choices are
not between $0$ and $k-1$).  Now we must similarly choose $\sigma_3$;
however, in this case we also have to choose it in such a way that the
sign $s(2)$ comes out correctly.  This constraint uniquely determines
$\sigma_3$, and in fact all the other $\sigma_m$ (for $m > 2$ and
$m\le 0$) are uniquely determined in this manner.

Thus, we have not only shown that the mapping is onto, but also that
each anyon model configuration has precisely $2k$ domain
configurations that map to it, for the $k$ choices of $\sigma_1$
above, and the two choices of $\sigma_2$.  Indeed, this $2k$-fold
degeneracy is easy to understand.  First of all, since the anyon
charges are functions of only the differences between the $\sigma_m$,
we can add a constant $c$ to all the $\sigma_m$ (mod) $k$ without
changing the anyon charges.  In fact, this is somewhat subtle.  For
example, if we have $\sigma_m < \sigma_{m+1}$, the corresponding anyon
charge is $f((\sigma_{m+1} - \sigma_m)/2)$.  Now, if we add a constant
$c$ such that $0 \leq \sigma_m+c < k$ but $\sigma_{m+1} + c \geq k$,
i.e., so that $\sigma_{m+1}$ cycles through, the new anyon type is
\be
\ba{c}
f( [(\sigma_m+c) - (\sigma_{m+1} + c - k) ] / 2) = \\
f(k/2 - (\sigma_{m+1} - \sigma_m)/2) = \\
 f((\sigma_{m+1} - \sigma_m)/2) \ea
\ee
by the definition of $f$.  Thus the anyon type still remains invariant.  The
signs of the domain model couplings also get modified under such a
cyclic shift of $\sigma_{m+1}$, but this is precisely canceled by the
corresponding modification of the $s'(m) =
(-1)^{(\sigma_{m-1}-\sigma_{m+1})} s(m)$ sign rule.  This cycling accounts for
a $k$-fold degeneracy; the factor of $2$ comes from the flip $\sigma_m
\rightarrow k-\sigma_m$, which preserves absolute values of
differences and signs of the couplings as well.  Thus the set of $2k$
pre-images of any anyon model configuration is simply an orbit of the
dihedral group ${\mathbb D}_k$, viewed as a subgroup of the symmetric group
$S_k$ acting on the configuration space of the Damle-Huse model.

Now that we understand the nature of the mapping in Eq. (\ref{fmap})
and (\ref{fmap2}), we can explicitly
verify that the inverse image of the fixed point ensemble (\ref{soln})
under this mapping is precisely the Damle-Huse fixed point. This
confirms that we
have a map that identifies the two fixed points and commutes with RG
evolution.  Furthermore, it means that the physical properties of the two systems
are the same, except those that might be affected by the $2k$ to $1$
nature of the mapping.  One of these is the existence of relevant
perturbations - we will prove that the anyon model has {\it no}
relevant perturbations, making it a stable phase.  

Before giving the formal argument regarding lack of relevant perturbations, we illustrate what happens with a physically
appealing picture.  Specifically, the domain model has $k-1$ relevant
perturbations \cite{DamleHuse} each of which can be described in terms
of one domain falling out of favor with respect to the rest (there is
a linear constraint since they can't all fall out of favor
simultaneously).  One might try to construct relevant perturbations of
the anyon model by mapping these relevant perturbations of the domain
model, as follows: each relevant perturbation can be thought of as a
functional on the configuration space of the domain model, so one can,
for a given anyon model configuration, sum up the values of the
relevant perturbation on all $2k$ of its pre-images (which are domain
model configurations).  This sum, however, turns out to be $0$, so no
relevant perturbation in the anyon model can be constructed this way.

Let us proceed by providing a formal proof based on symmetry.
The proof is by contradiction.  Supposing we had a relevant
perturbation of the anyon model,  we could then pull it back to a
relevant perturbation of the domain model (given a map $M:
X\rightarrow Y$ of spaces, the pullback map on function spaces
$F(Y)\rightarrow F(X)$ is defined by $f \rightarrow g$ where $g(x) =
f(M(x))$).  Because of the $2k$ to $1$ nature of the mapping, this
would yield a ${\mathbb D}_k$ symmetric relevant perturbation of the domain
model; in particular it would also be ${\mathbb Z}_k \subset {\mathbb D}_k$
symmetric.  The $k-1$ relevant perturbations of the domain
model, however, are not ${\mathbb Z}_k$ symmetric, because they are described in
terms of one of the $k$ domains falling out of favor with respect to
the others.  Indeed, they form a $k-1$ dimensional non-trivial
irreducible representation of ${\mathbb Z}_k$.  Our putative relevant
deformation is ${\mathbb Z}_k$ symmetric, i.e., lies in the trivial
representation of ${\mathbb Z}_k$.  Thus we have found a non-existent
relevant perturbation of the domain model, a contradiction.  Of
course, this analysis does not include possible perturbations by the
addition of interactions $\Q^{(f)}$ with $f>1$, but we have
already shown that the Hamiltonian is stable with respect to such
perturbations in section III, i.e., we showed they are irrelevant.  We
have therefore realized all of the odd $k$ $S_k$ symmetric
multi-critical points as stable phases in the $SU(2)_k$ anyon chains.

\section{Conclusions}

In this paper we analyzed in full generality spin-chains made of
non-abelian quasiparticles arising in the tensor categories of
$SU(2)_k$ for odd $k$, in the limit of strong randomness. We have found a realization of the odd $k$ $S_k$ symmetric infinite
randomness multi-critical fixed points of Damle and Huse \cite{DamleHuse} as
critical {\it stable phases} of disordered $SU(2)_k$ anyon chains.  We
have shown that the $SU(2)_k$ fixed point is stable by analyzing the
RG flow equations around it, and also by explicit mapping to the
Damle-Huse model.  Key in our analysis was our use of a basis of
interaction operators $\Q^(f)$ that behave well with respect to
real-space decimations, i.e., operators $\Q^(f)$ that have well defined scaling under
the RG.  We found that $\Q^(1)$ is the most relevant operator, and thus
the only one appearing in the effective Hamiltonian (\ref{hanyon1}) at
the fixed point.  This effective elimination of all but one
interaction operator is what resolves the a priori problem of having a
potential multitude of energy scales associated with the multiple
fusion products of each neighboring pair of anyons.  Indeed, at the
fixed point (\ref{hanyon1}) each bond is characterized by only one
energy scale: the coefficient in front of the $\Q^(1)$ operator.

Recall that the motivation for our study was the search for new
universality classes of infinite randomness fixed points. In that
sense, our analysis led to a disappointment: The $SU(2)_k$ anyonic
chains exhibited the same behavior as random spin-$(k-1)/2$ chains, as
though the differences between the extremely distinct Hilbert spaces
of the two systems were essentially washed out in the
strong randomness limit, leading to the same infinite-randomness fixed
points. Nevertheless, a crucial difference arose: the permutation
symmetric fixed points mark {\it stable phases} of the $SU(2)_k$
random spin chains, as opposed to unstable points in the ordinary spin
chains. 

One natural question is then whether $SU(N)_k$ for higher $N$
behaves any differently.  It is plausible that using an approach
similar to the one in this paper, with the relevant interactions as
exchanges of a certain anyon type, will yield the already known fixed
points, essentially because the charge of the exchanged anyon will
pick out a preferred $SU(2) \subset SU(N)$ and will decompose the
problem to the $SU(2)$ cases already studied.  This does not, of
course, rule out the possibility of more symmetric and exotic fixed
points, which may arise with some fine tuning, for instance.

%First of
%all, we have not discussed the case of even $k$; it is plausible that
%it works out similarly, but with the mapping (\ref{fmap}) modified to
%accomodate the different folding of even $k$ topological charges.  Also, from the picture we have constructed,
%one can study many possible flows
%among the associated fixed points in the infinite randomness limit.  In addition to the flow from the
%random singlet to any of these points (a generalization of the flow
%constructed in [\onlinecite{us}]), we can naturally embed the
%$SU(2)_k$ fixed point as an unstable critical point in the
%$SU(2)_{nk}$ anyon chain system, and obtain a flow from one to the
%other.

Another line of investigation deals with the relation to the uniform
$SU(2)_k$ anyonic chains of \onlinecite{Qpeople3}.  There, a
"topological" symmetry stabilizes an otherwise $k$ multi-critical
point CFT low energy spectrum.  A physical interpretation of this
phenomenon is given in terms of separating out the left and right
moving modes of the chain, while creating a different topological
liquid between them - the topological symmetry then eliminates
relevant tunneling operators between the two modes.  One could then
add some very weak disorder to this system: at short distances
the picture of two modes separated by a topological liquid is
preserved, while at long distances the disorder grows and the dynamics
is controlled by the infinite randomness fixed point discussed in this
paper.  It is interesting to try to find a physical picture for the
infinite randomness phase - perhaps with the intervening liquid having broken
up into disconnected islands - which may also yield the stability
argument of the fixed points we found. 

On the other hand, a picture that is similar to the one of separating the right and
left moving modes away from each other but which also applies to the strongly
disordered system may provide clues to the understanding of the behavior of non-abelian anyons
interacting on random planar graphs. Indeed, it would be interesting to see if any of the ideas developed in this
paper have application to, say, a two (or higher) dimensional disordered lattice
of anyons.  So far we haven't made progress in this direction.

\acknowledgments
We would like to thank John Preskill and Simon Trebst for useful
discussions.  Also, we would especially like to thank David Huse for
useful discussions during the early part of this work.  H-H.L. and
P.T. were supported by the Summer Undergraduate Research Fellowship at the California Institute of Technology.
L.F. and G.R. would like to acknowledge support from the Institute for
Quantum Information under NSF grants PHY-0456720 and PHY-0803371, and
from the Packard Foundation.

\section{Appendix}

In this appendix, we derive equation (\ref{apeq}) for the energy spacing between the different fusion channels in the $\Q^{(1)}$ interaction, show that $\Q^{(1)}$ is the most relevant operator for both first and second order decimations, and derive the sign rules for the changes of the sign of the couplings under first order decimations.  For the analysis we will need the following expression \cite{slingerland} for the $F$-matrix (or $6$-$j$ symbols) of $SU(2)_k$:

\begin{widetext}
\begin{equation}
\label{sixjsymbs}
\begin{array}{l}
\left( {F \,}_{j_1 j_2}^{j j_3} \right)_{j_{12}}^{j_{23}} \rule[-7mm]{0mm}{5mm}=\\[2mm]
~~~~~
\sqrt{\qnr{2j_{12}+1}\qnr{2j_{23}+1}}\,
\Delta(j_1,j_2,j_{12})\Delta(j_{12},j_3,j)
\Delta(j_2,j_3,j_{23})\Delta(j_1,j_{23},j)  \\[2mm]
~~~~~
\times \sum_{z} \left\{
\frac{(-1)^{z}\qnr{z+1}!}{\qnr{z-j_1-j_2-j_{12}}!
\qnr{z-j_{12}-j_3-j}!\qnr{z-j_2-j_3-j_{23}}!\qnr{z-j_1-j_{23}-j}!}
\right.
 \\[2mm]
~~~~~~~~~~~~~~~~~
\times \left. 
\frac{1}{\qnr{j_1+j_2+j_3+j-z}!\qnr{j_1+j_{12}+j_3+j_{23}-z}!
\qnr{j_2+j_{12}+j+j_{23}-z}!}
\right\}
\end{array} 
\end{equation}
\end{widetext}Here the q-numbers are defined as in (\ref{qnreq}).  The
sum is over all $z$ for which all the $q$-factorials are well-defined,
i.e., such that the arguments are $\geq 0$, and
\begin{equation}
\Delta(a,b,c):=
\sqrt{\frac{\qnr{-a+b+c}!\qnr{a-b+c}!\qnr{a+b-c}!}{\qnr{a+b+c+1}!}}.
\end{equation}

Let us first analyze the operator $\Q_{p,p+1}^{(1)}$.  To simplify notation, we denote $\{ i_p, i_{p+1} \}$ by $\{ i, j \}$ and drop the subscripts on $\Q^{(\beta)}$ and the projectors $\P^{(\beta)}$.  Applying an $F$-matrix move and noting the normalization on the projector $\P_{p,p+1}^{(f)}$, we obtain

\be
\Q^{(1)} = \sum_{f=|i - j|}^{\text{min }(i+j,k-i-j)} \frac{(F_{i i}^{j j} )_1^f} {(F_{i i}^{j j})_0^f} \, \P^{(f)}
\ee We would like to know the $f$-dependence of the coefficient in front of $\P^{(f)}$.  Plugging into (\ref{sixjsymbs}), up to an $f$-independent prefactor, the coefficient is

\be
\left( \qnr{f}^2 + \qnr{f+1}^2 - \qnr{|i - j|}^2 - \qnr{i+j + 1}^2 \right)
\label{q1}
\ee giving us (\ref{apeq}) as desired.  Note that the quantity in the brackets is an increasing function of $f$.

Let us now work out the decimation rules for the Hamiltonian (\ref{hanyon1}), and in particular show that the $\Q$ operators are eigenvectors with respect to decimations, with $\Q^{(1)}$ having the largest eigenvalue.  Pick the largest coupling $J_{p,p+1}$.  According to (\ref{q1}), anyons $p$ and $p+1$ will be fused into an anyon of total topological charge $| i_p - i_{p+1} |$ for $J_{p,p+1}>0$ and into an anyon of charge $\text{max } (i_p + i_{p+1}, k - i_p - i_{p+1})$ for $J_{p,p+1}<0$.  Suppose first that this composite anyon charge is nonzero.  In this case we need to use first order perturbation theory to work out the effective coupling of this composite anyon to its neighbors.  For simplicity we only consider $i_{p+2}$, the neighbor to the right, and to simplify notation we denote $i_p, i_{p+1}, i_{p+2}$ by $i,j,m$.  

Let's first deal with a specific case, say having $i>j$ fuse to $i-j$.  According to first order perturbation theory, the effective coupling between $i-j$ and $m$ is given by

\be
\P_{p, p+1} \, \Q_{p+1,p+2}^{(1)} \, \P_{p,p+1}
\ee Graphically we can see (figure \ref{acf5}) that this effective coupling is still a multiple of $\Q^{(1)}$.  There is a finite factor in front of the coefficient, as well as a possible sign, but we already see that the form (\ref{hanyon1}) is preserved by first order decimations.  In fact, it will be useful to go a little further and examine the effect of such a decimation on the other interactions $\Q_{p+1,p+2}^{(\beta)}$.  Again, we graphically see that the effective coupling will be a constant multiple of $\Q_{p+1,p+2}^{(\beta)}$.  We are only interested in the sign and $\beta$ dependence of this constant, which turns out to be equal to the constant $C$ in figure \ref{acf5}, this being the evaluation of the graph in the dashed box.  This is

\be \label{mueq}
\mu(\beta)=\frac{(F_{\alpha j}^{\alpha j})_i^\beta}{(F_{jj}^{jj})_0^\beta}
\ee where $\alpha$ is the fusion product of $i$ and $j$, in this case $\alpha = i-j$.  Plugging into (\ref{sixjsymbs}), we see that the $\beta$ dependence of (\ref{mueq}) is

\begin{widetext}
\be
\mu(\beta) = (-1)^\beta \left( \qnr{2j-\beta}! \qnr{2i - 2j + \beta+1}! \qnr{2j+\beta+1}! \qnr{2i - 2j -\beta}! \right)^{-1/2}
\ee We observe that the magnitude of this quantity decreases as a function of $\beta$.  To see this, note that

\be
\mu(\beta) / \mu(\beta-1) = - \left(\frac{\qnr{2j-\beta+1}}{\qnr{2j+\beta+1}}\right)^{1/2} \left(\frac{\qnr{2i-2j-\beta+1}}{\qnr{2i-2j+\beta+1}}\right)^{1/2} = -h(j) \, h(i-j)
\ee
\end{widetext} It's easy to see using the explicit expression for the $q$-numbers in terms of roots of unity that $h(j) = h(k/2-j)^{-1}$ and that $h(j)$ is an increasing function of $j$.  Therefore 

\be h(j) \, h(i-j) \leq h(j) \, h(k/2 - j) = 1 \ee Thus we have shown that $\Q^{(1)}$ has the highest eigenvalue under first order decimations.  We've only considered the case $i>j$, $\alpha = i-j$, but similar arguments apply to other cases (for the case of $\alpha = k-i-j$ we need to use the symmetries of the $F$-matrix discussed in [\onlinecite{slingerlandthesis}]) to show that $\mu(\beta)$ decreases as a function of $\beta$. 

The one thing we need to know explicitly for the mapping to the Damle-Huse
model is whether first order decimations flip the sign of the
neighboring couplings, when those are of the form $\Q^{(1)}$.  For the
case just considered, $i>j, \alpha=i-j$, the sign is $(-1)^\beta=-1$.
For the rest of the cases the sign can be read off from the factors of
$(-1)^z$ in (\ref{sixjsymbs}): when $i<j,\, \alpha=j-i$, the sign is
$+1$; when $\alpha=i+j<k/2$, the sign is $+1$; and when $\alpha =
k-i-j<k/2$ the sign is $-1$.

\begin{figure}
\includegraphics[width=6cm]{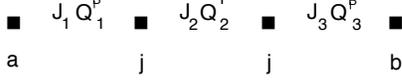}
\caption{The setting for second order decimations.  The anti-ferromagnetic coupling $J_2$ is much larger than $J_1$ and $J_3$, causing the two middle anyons to combine into a charge singlet. \label{acf6}}
\end{figure}

\begin{figure}
\includegraphics[width=3.5cm]{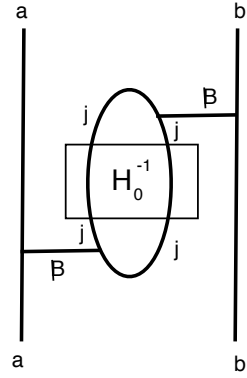}
\caption{Graphical expression for the effective interaction between $a$ and $b$ generated at second order in perturbation theory.  Here $H_0$ is as defined in the text.  Its inverse is computed and expressed in graphical form by expanding in projection operators.  This is one of the two terms that contribute to an energy splitting between the fusion products of $a$ and $b$.  Note that both the original interaction with $a$ and $b$ exchange the same topological charge $\beta$ - otherwise the second order correction is zero by topological charge conservation, as is clear from the figure. \label{acf7}}
\end{figure}

Consider now the second order decimations.  The setup here is that we have four consecutive anyons, with the middle two anti-ferromagnetically fusing to the trivial channel, so we can label their topological charges $a, j, j, b$.  The picture is as in figure \ref{acf6}.  The bare Hamiltonian is

\be H_0 = J_2 (\Q_2^{(1)} - < \Q_2^{(1)} >) \ee where $<\Q_2^{(1)}>$ is the expectation value of $\Q_2^{(1)}$ in the state where the fusion product of the two $j$'s is trivial.  The interaction Hamiltonian is

\be H' = J_1 \Q_1^{\beta} + J_3 \Q_3^{\beta} \ee Note that the two interactions must have the same $\beta$ by charge conservation (see figure \ref{acf7}).  The induced effective Hamiltonian between $a$ and $b$ at second order is

\be \He= \P \, H' \, H_0^{-1} H' \ee  $\He$ acts on the subspace where the two middle anyons, both of charge $j$, fuse to the trivial channel.  The inverse $H_0^{-1}$ is well defined in this expression because $H'$ acting on this subspace gives a vector orthogonal to the subspace.  We have, up to a multiple of the identity operator,

\begin{eqnarray}
\He&=& \frac{J_1 J_3}{J_2} \P \, \Q_1^{\beta} (\Q_2^{(1)} - <\Q_2^{(1)}>)^{-1} \Q_3^{\beta} \nonumber \\ &+& \frac{J_1 J_3}{J_2} \P \, \Q_3^{\beta} (\Q_2^{(1)} - <\Q_2^{(1)}>)^{-1} \Q_1^{\beta} \end{eqnarray} Using the $F$-matrix we calculate that

\be (\Q_2^{(1)} - <\Q_2^{(1)}>)^{-1} = \sum_\gamma \left( \frac{(F_{jj}^{jj})_1^\gamma}{(F_{jj}^{jj})_0^\gamma} - \frac{(F_{jj}^{jj})_0^1}{(F_{jj}^{jj})_0^0} \right)^{-1} \, \P^{(\gamma)} \label{qinv} \ee From figure \ref{acf6} we then see that the effective operator between $a$ and $b$ is

\be c \, \left( \frac{(F_{jj}^{jj})_1^\beta}{(F_{jj}^{jj})_0^\beta} - \frac{(F_{jj}^{jj})_0^1}{(F_{jj}^{jj})_0^0} \right)^{-1} \, \Q^{(\beta)} \ee where $c$ is a constant containing the normalization of the projection operator $\P^{(\beta)}$ relative to its graphical representation, times the numerical factor one gets from reducing the portion of the graph in figure \ref{acf7} between the two $\beta$'s using $F$-matrix moves.  The product in front of $\Q^{(\beta)}$ comes out to

\be \left( (F_{jj}^{jj})_1^\beta - \frac{(F_{jj}^{jj})_0^1}{(F_{jj}^{jj})_0^0} \, (F_{jj}^{jj})_0^\beta \right)^{-1} \ee  Evaluating this expression using (\ref{sixjsymbs}) we see that, up to $\beta$-independent factors, it is equal to

%\be \frac{ (-1)^\beta \, \qnr{2j} \, \qnr{2j+1}\,
 % \qnr{2j+2}}{\qnr{2\beta+1}^{1/2} \, \qnr{3}^{1/2}
 % \left(\qnr{\beta}^2 + \qnr{\beta+1}^2 - 1\right)} \ee
 
 \be (-1)^\beta \, \qnr{2\beta+1}^{-1/2} \, \left(\qnr{\beta}^2 + \qnr{\beta+1}^2 - 1 \right)^{-1} \ee Though this is
not a decreasing function of $\beta$, both of the factors are minimized in absolute value at $\beta$=1, so the absolute value of the
expression is maximized at $\beta=1$.  Thus we have shown that second
order decimations also have the $\Q^{(\beta)}$ as eigenvectors, and
$\Q^{(1)}$ has the highest magnitude eigenvalue. The second order decimation
  rule for $\beta=1$ is then:
\be
\He=\frac{J_1 J_3}{J_2} C_j \Q^{(1)}_{1,4},
\label{2ndDEC}
\ee where

\be \label{cj}
C_j = \frac{\qnr{2}^2}{\qnr{3}} \left( \qnr{2j+1} - \qnr{2j+1}^{-1} \right)^{-1}.
\ee

\bibliography{anyonchainrefs}

\begin{thebibliography}{17}
\expandafter\ifx\csname natexlab\endcsname\relax\def\natexlab#1{#1}\fi
\expandafter\ifx\csname bibnamefont\endcsname\relax
  \def\bibnamefont#1{#1}\fi
\expandafter\ifx\csname bibfnamefont\endcsname\relax
  \def\bibfnamefont#1{#1}\fi
\expandafter\ifx\csname citenamefont\endcsname\relax
  \def\citenamefont#1{#1}\fi
\expandafter\ifx\csname url\endcsname\relax
  \def\url#1{\texttt{#1}}\fi
\expandafter\ifx\csname urlprefix\endcsname\relax\def\urlprefix{URL }\fi
\providecommand{\bibinfo}[2]{#2}
\providecommand{\eprint}[2][]{\url{#2}}

\bibitem[{\citenamefont{Kitaev}(2003)}]{kitaev}
\bibinfo{author}{\bibfnamefont{A.~Y.} \bibnamefont{Kitaev}},
  \bibinfo{journal}{Annals Phys.} \textbf{\bibinfo{volume}{303}},
  \bibinfo{pages}{2} (\bibinfo{year}{2003}),
  \urlprefix\url{http://www.citebase.org/abstract?id=oai:arXiv.org:quant-ph/97%
07021}.

\bibitem[{\citenamefont{Collins}(2006)}]{collins}
\bibinfo{author}{\bibfnamefont{G.~P.} \bibnamefont{Collins}},
  \bibinfo{journal}{Scientific American} \textbf{\bibinfo{volume}{4}},
  \bibinfo{pages}{57} (\bibinfo{year}{2006}).

\bibitem[{\citenamefont{Feiguin et~al.}(2007)\citenamefont{Feiguin, Trebst,
  Ludwig, Troyer, Kitaev, Wang, and Freedman}}]{Qpeople1}
\bibinfo{author}{\bibfnamefont{A.}~\bibnamefont{Feiguin}},
  \bibinfo{author}{\bibfnamefont{S.}~\bibnamefont{Trebst}},
  \bibinfo{author}{\bibfnamefont{A.~W.~W.} \bibnamefont{Ludwig}},
  \bibinfo{author}{\bibfnamefont{M.}~\bibnamefont{Troyer}},
  \bibinfo{author}{\bibfnamefont{A.}~\bibnamefont{Kitaev}},
  \bibinfo{author}{\bibfnamefont{Z.}~\bibnamefont{Wang}}, \bibnamefont{and}
  \bibinfo{author}{\bibfnamefont{M.~H.} \bibnamefont{Freedman}},
  \bibinfo{journal}{Phys. Rev. Lett.} \textbf{\bibinfo{volume}{98}},
  \bibinfo{pages}{160409} (\bibinfo{year}{2007}).

\bibitem[{\citenamefont{Trebst et~al.}(2008)\citenamefont{Trebst, Ardonne,
  Feiguin, Huse, Ludwig, and Troyer}}]{Qpeople2}
\bibinfo{author}{\bibfnamefont{S.}~\bibnamefont{Trebst}},
  \bibinfo{author}{\bibfnamefont{E.}~\bibnamefont{Ardonne}},
  \bibinfo{author}{\bibfnamefont{A.}~\bibnamefont{Feiguin}},
  \bibinfo{author}{\bibfnamefont{D.~A.} \bibnamefont{Huse}},
  \bibinfo{author}{\bibfnamefont{A.~W.~W.} \bibnamefont{Ludwig}},
  \bibnamefont{and} \bibinfo{author}{\bibfnamefont{M.}~\bibnamefont{Troyer}},
  \bibinfo{journal}{Physical Review Letters} \textbf{\bibinfo{volume}{101}},
  \bibinfo{pages}{050401} (\bibinfo{year}{2008}),
  \urlprefix\url{doi:10.1103/PhysRevLett.101.050401}.

\bibitem[{\citenamefont{Fisher}(1994)}]{DSF94}
\bibinfo{author}{\bibfnamefont{D.~S.} \bibnamefont{Fisher}},
  \bibinfo{journal}{Phys. Rev. B} \textbf{\bibinfo{volume}{50}},
  \bibinfo{pages}{3799} (\bibinfo{year}{1994}).

\bibitem[{\citenamefont{Fisher}(1995)}]{DSF95}
\bibinfo{author}{\bibfnamefont{D.~S.} \bibnamefont{Fisher}},
  \bibinfo{journal}{Phys. Rev. B} \textbf{\bibinfo{volume}{51}},
  \bibinfo{pages}{6411} (\bibinfo{year}{1995}).

\bibitem[{\citenamefont{Damle and Huse}(2002)}]{DamleHuse}
\bibinfo{author}{\bibfnamefont{K.}~\bibnamefont{Damle}} \bibnamefont{and}
  \bibinfo{author}{\bibfnamefont{D.~A.} \bibnamefont{Huse}},
  \bibinfo{journal}{Phys. Rev. Lett.} \textbf{\bibinfo{volume}{89}},
  \bibinfo{pages}{277203} (\bibinfo{year}{2002}).

\bibitem[{\citenamefont{Monthus et~al.}(1998)\citenamefont{Monthus, Golinnelli,
  and Jolicoeur}}]{MGJ}
\bibinfo{author}{\bibfnamefont{C.}~\bibnamefont{Monthus}},
  \bibinfo{author}{\bibfnamefont{O.}~\bibnamefont{Golinnelli}},
  \bibnamefont{and}
  \bibinfo{author}{\bibfnamefont{T.}~\bibnamefont{Jolicoeur}},
  \bibinfo{journal}{Phys. Rev. B} \textbf{\bibinfo{volume}{58}},
  \bibinfo{pages}{805} (\bibinfo{year}{1998}).

\bibitem[{\citenamefont{Hyman and Yang}(1997)}]{HymanYang}
\bibinfo{author}{\bibfnamefont{R.~A.} \bibnamefont{Hyman}} \bibnamefont{and}
  \bibinfo{author}{\bibfnamefont{K.}~\bibnamefont{Yang}},
  \bibinfo{journal}{Phys. Rev. Lett.} \textbf{\bibinfo{volume}{78}},
  \bibinfo{pages}{1783} (\bibinfo{year}{1997}).

\bibitem[{\citenamefont{Refael et~al.}(2002)\citenamefont{Refael, Kehrein, and
  Fisher}}]{RKF}
\bibinfo{author}{\bibfnamefont{G.}~\bibnamefont{Refael}},
  \bibinfo{author}{\bibfnamefont{S.}~\bibnamefont{Kehrein}}, \bibnamefont{and}
  \bibinfo{author}{\bibfnamefont{D.~S.} \bibnamefont{Fisher}},
  \bibinfo{journal}{Phys. Rev. B} \textbf{\bibinfo{volume}{66}},
  \bibinfo{pages}{060402} (\bibinfo{year}{2002}).

\bibitem[{\citenamefont{Bonesteel and Yang}(2007)}]{YB}
\bibinfo{author}{\bibfnamefont{N.~E.} \bibnamefont{Bonesteel}}
  \bibnamefont{and} \bibinfo{author}{\bibfnamefont{K.}~\bibnamefont{Yang}},
  \emph{\bibinfo{title}{Infinite-randomness fixed points for chains of
  non-abelian quasiparticles}} (\bibinfo{year}{2007}).

\bibitem[{\citenamefont{Fidkowski et~al.}(2008)\citenamefont{Fidkowski, Refael,
  Bonesteel, and Moore}}]{us}
\bibinfo{author}{\bibfnamefont{L.}~\bibnamefont{Fidkowski}},
  \bibinfo{author}{\bibfnamefont{G.}~\bibnamefont{Refael}},
  \bibinfo{author}{\bibfnamefont{N.}~\bibnamefont{Bonesteel}},
  \bibnamefont{and} \bibinfo{author}{\bibfnamefont{J.}~\bibnamefont{Moore}},
  \emph{\bibinfo{title}{Infinite randomness phases and entanglement entropy of
  the disordered golden chain}} (\bibinfo{year}{2008}),
  \urlprefix\url{http://www.citebase.org/abstract?id=oai:arXiv.org:0807.1123}.

\bibitem[{\citenamefont{Gils et~al.}(2008)\citenamefont{Gils, Ardonne, Trebst,
  Ludwig, Troyer, and Wang}}]{Qpeople3}
\bibinfo{author}{\bibfnamefont{C.}~\bibnamefont{Gils}},
  \bibinfo{author}{\bibfnamefont{E.}~\bibnamefont{Ardonne}},
  \bibinfo{author}{\bibfnamefont{S.}~\bibnamefont{Trebst}},
  \bibinfo{author}{\bibfnamefont{A.~W.~W.} \bibnamefont{Ludwig}},
  \bibinfo{author}{\bibfnamefont{M.}~\bibnamefont{Troyer}}, \bibnamefont{and}
  \bibinfo{author}{\bibfnamefont{Z.}~\bibnamefont{Wang}},
  \emph{\bibinfo{title}{Topological stability of anyonic quantum spin chains
  and formation of new topological liquids}} (\bibinfo{year}{2008}),
  \urlprefix\url{http://arxiv.org/abs/0810.2277}.

\bibitem[{\citenamefont{Ma et~al.}(1979)\citenamefont{Ma, Dasgupta, and
  Hu}}]{MaDas1979}
\bibinfo{author}{\bibfnamefont{S.~K.} \bibnamefont{Ma}},
  \bibinfo{author}{\bibfnamefont{C.}~\bibnamefont{Dasgupta}}, \bibnamefont{and}
  \bibinfo{author}{\bibfnamefont{C.~K.} \bibnamefont{Hu}},
  \bibinfo{journal}{Phys. Rev. Lett.} \textbf{\bibinfo{volume}{43}},
  \bibinfo{pages}{1434} (\bibinfo{year}{1979}).

\bibitem[{\citenamefont{Dasgupta and Ma}(1980)}]{MaDas1980}
\bibinfo{author}{\bibfnamefont{C.}~\bibnamefont{Dasgupta}} \bibnamefont{and}
  \bibinfo{author}{\bibfnamefont{S.~K.} \bibnamefont{Ma}},
  \bibinfo{journal}{Phys. Rev. B} \textbf{\bibinfo{volume}{22}},
  \bibinfo{pages}{1305} (\bibinfo{year}{1980}).

\bibitem[{\citenamefont{Slingerland and Bais}(2001)}]{slingerland}
\bibinfo{author}{\bibfnamefont{J.~K.} \bibnamefont{Slingerland}}
  \bibnamefont{and} \bibinfo{author}{\bibfnamefont{F.~A.} \bibnamefont{Bais}},
  \bibinfo{journal}{Nuclear Physics B} \textbf{\bibinfo{volume}{612}},
  \bibinfo{pages}{229} (\bibinfo{year}{2001}),
  \urlprefix\url{http://www.citebase.org/abstract?id=oai:arXiv.org:cond-mat/01%
04035}.

\bibitem[{\citenamefont{Slingerland}()}]{slingerlandthesis}
\bibinfo{author}{\bibfnamefont{J.~K.} \bibnamefont{Slingerland}},
  \emph{\bibinfo{title}{Hopf symmetry and its breaking; braid statistics and
  confinement in planar physics}},
  \urlprefix\url{http://www.stp.dias.ie/~slingerland/thesis.pdf}.

\end{thebibliography}

\end{document}